\documentclass[twocolumn,aps,prl]{revtex4}
\usepackage{amsmath}
\usepackage{graphicx}
\usepackage{dcolumn}
\usepackage{bm}

\preprint{}
\input{tcilatex}

\begin{document}

\title{Electromagnetic Wave Propagation in Media with Indefinite
Permittivity and Permeability Tensors}
\author{D. R. Smith and D. Schurig}
\date{\today }

\begin{abstract}
We study the behavior of wave propagation in materials for which not all of
the principle elements of the permeability and permittivity tensors have the
same sign. We find that a wide variety of effects can be realized in such
media, including negative refraction, near-field focusing and high impedance
surface reflection. In particular a bi-layer of these materials can transfer
a field distribution from one side to the other, including near-fields,
without requiring internal exponentially growing waves.
\end{abstract}

\maketitle
\affiliation{Physics Department, University of California, San Diego, La Jolla, CA, 92093}

The range of available electromagnetic material properties has been
broadened by recent developments in structured media, notably Photonic Band
Gap materials and metamaterials. These media have allowed the realization of
solutions to Maxwell's equations not available in naturally occurring
materials, fueling the discovery of new physical phenomena and the
development of devices. Photonic crystals, for example, can modify the
radiative density of states associated with nearby electromagnetic sources.
Optical effects such as superradiance \cite{John+Quang}, enhanced or
inhibited spontaneous emission \cite{Koenderink}, ultrarefraction \cite%
{Kosaka} and even negative refraction \cite{notomi,gralak} have been
predicted for various photonic lattice configurations. \ 

Photonic crystal effects typically occur when the wavelength is on the same
order or smaller than the lattice constant of the crystal. \ Metamaterials,
on the other hand, have unit cell dimensions much smaller than the
wavelength of interest. A homogenization process can thus be applied,
allowing the otherwise complicated composite medium to be described
conveniently by a permittivity tensor ($\mathbf{\varepsilon }$) and a
permeability tensor ($\mathbf{\mu }$) \cite{Halevi}, rather than by band
diagrams.

In 2000, it was shown experimentally that a metamaterial composed of
periodically positioned scattering elements, all conductors, could be
interpreted as having simultaneously a negative effective $\varepsilon $ and
a negative effective $\mu $ \cite{smithPRL}. Such a medium had been
previously shown by Veselago to be consistent with Maxwell's equations \cite%
{veselago}, but had never been demonstrated in a naturally occurring
material or compound. A medium with simultaneously isotropic and negative $%
\varepsilon $ and $\mu $ supports propagating solutions whose phase and
group velocities are antiparallel; equivalently, such a material can be
rigorously described as having a negative index of refraction\cite%
{smithkroll}. An experimental observation of negative refraction was
reported using a metamaterial composed of wires and split ring resonators
deposited lithographically on circuit board material \cite{shelby}.

The prospect of negative refractive materials has generated considerable
interest, as this simply stated material condition suggests the possibility
of extraordinary wave propagation phenomena, including near-field focusing %
\cite{perfectLens}. So remarkable have been the claims surrounding negative
refraction, that some researchers have been prompted to examine critically
the achievability of negative refraction in existing metamaterials \cite%
{Valanju, insaneGarcia}. While such concerns might appear relevant in the
context of frequency-dispersive materials, the interpretation of these
structured materials as negative refractive has been entirely consistent
with all aspects of reported experimental data \cite{smithPRL,shelby}, as
well as with numerical simulations of both monochromatic and modulated beams %
\cite{kongBeam,modulate}. \ For the purposes of this Letter, we thus assume
that the descriptions presented here will be applicable to real materials,
although the extent to which this is true remains a topic of active pursuit.

Lindell et al. \cite{lindell} have shown that the property of negative
refraction is not confined to materials with negative definite $\epsilon $
and $\mu $, but can be expected to occur in certain classes of uniaxially
anisotropic media. In addition to exhibiting negative refraction, some
classes of anisotropic media have very unusual dispersion relations,
characterized for example by hyperbolic dispersion curves. The range of
electromagnetic wave propagation behavior expected for such materials
greatly extends beyond that available in isotropic negative refractive
materials, and motivates our pursuit here to find configurations that take
advantage of the unique properties.

\begin{figure}[tbp]
\includegraphics[ width=2.73in, height=3.84in ]{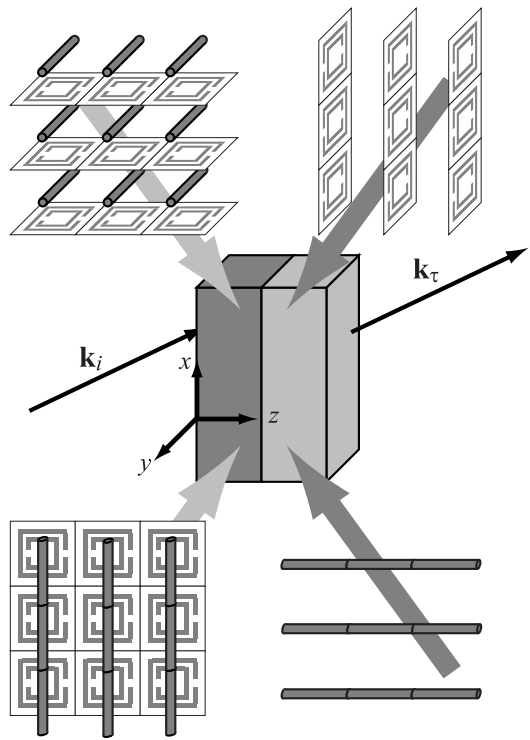}
\caption{A bilayer composed of indefinite media implemented using split ring
resonators and straight wires. \ The structures in the top of the figure
implement \emph{never cutoff} media for electric $y$-polarization. \ Top
left is negative refracting and top right is positve refracting. \ The
structures in the bottom of the figure do the same for magnetic $y$%
-polarization.}
\label{implementation fig}
\end{figure}

To simplify the proceeding analysis, we assume a linear material with no
magnetoelectric coupling, so that the media can be fully described by a
permittivity, $\mathbf{\varepsilon }$, and permeability, $\mathbf{\mu }$,
tensor. [Bianisotropy effects have typically played a minor role in the
overall response of the experimental metamaterials, and can be mitigated by
design \cite{Marques}.) We further assume that these tensors are
simultaneously diagonalizable, having the form 
\begin{equation}
\mathbf{\varepsilon }=\left( 
\begin{array}{ccc}
\varepsilon _{x} & 0 & 0 \\ 
0 & \varepsilon _{y} & 0 \\ 
0 & 0 & \varepsilon _{z}%
\end{array}%
\right) \;\mathbf{\mu }=\left( 
\begin{array}{ccc}
\mu _{x} & 0 & 0 \\ 
0 & \mu _{y} & 0 \\ 
0 & 0 & \mu _{z}%
\end{array}%
\right) .
\end{equation}%
Metamaterials can be readily constructed that closely approximate these $%
\mathbf{\varepsilon }$ and $\mathbf{\mu }$ tensors, with elements of either
algebraic sign. In fact, the scattering elements comprising the
metamaterials used to demonstrate negative refraction \cite{shelby} are
appropriate building blocks for the classes of materials to be discussed
here, (see Fig. \ref{implementation fig}.)

We are interested in an anisotropic medium in which not all of the principle
components of the $\mathbf{\varepsilon }$ and $\mathbf{\mu }$ tensors have
the same sign. For brevity, we refer to such a medium as \emph{indefinite}.
\ We will consider layered media with surfaces normal to one of the
principle axes, which we define to be the $z$-axis. We demonstrate our
analysis using a plane wave with the electric field polarized along the $y$%
-axis, though it is generally possible to construct media that are
polarization independent, or exhibit different classes of behavior for
different polarizations. 
\begin{equation}
\mathbf{E}=\widehat{\mathbf{y}}e^{i\left( k_{x}x+k_{z}z-\omega t\right) }.
\end{equation}%
Plane wave solutions to Maxwell's equations with this polarization have $%
k_{y}=0$ and satisfy: 
\begin{equation}
k_{z}^{2}=\varepsilon _{y}\mu _{x}\frac{\omega ^{2}}{c^{2}}-\frac{\mu _{x}}{%
\mu _{z}}k_{x}^{2}.  \label{dispersion}
\end{equation}%
Since we have no $x$ or $y$ oriented boundaries or interfaces, real
exponential solutions, which result in field divergence when unbounded, are
not allowed in those directions. Thus $k_{x}$ is restricted to be real.
Also, since $k_{x}$ represents a variation transverse to the surfaces of our
layered media, it is conserved across the layers, and naturally
parameterizes the solutions.

In the absence of losses, the sign of $k_{z}^{2}$ can be used to distinguish
the nature of the plane wave solutions. $k_{z}^{2}>0$\ corresponds to real
valued $k_{z}$ and propagating solutions. $k_{z}^{2}<0$ corresponds to
imaginary $k_{z}$ and exponentially growing or decaying (evanescent)
solutions. When $\varepsilon _{y}\mu _{z}>0$, there will be a value of $%
k_{x} $ for which $k_{z}^{2}=0$. This value, which we denote $k_{c}$, is the
cutoff wave vector separating propagating from evanescent solutions. From
Eq. (\ref{dispersion}) this value is $k_{c}=\frac{\omega }{c}\sqrt{%
\varepsilon _{y}\mu _{z}}.$ We identify four classes of media based on their
cutoff properties:

\begin{equation*}
\begin{tabular}{|l|l|l|l|}
\hline
& \multicolumn{2}{|l|}{\textbf{Media Conditions}} & \textbf{Propagation} \\ 
\hline
Cutoff & $\varepsilon _{y}\mu _{x}>0$ & $\mu _{x}/\mu _{z}>0$ & $k_{x}<k_{c}$
\\ \hline
Anti-Cutoff & $\varepsilon _{y}\mu _{x}<0$ & $\mu _{x}/\mu _{z}<0$ & $%
k_{x}>k_{c}$ \\ \hline
Never Cutoff & $\varepsilon _{y}\mu _{x}>0$ & $\mu _{x}/\mu _{z}<0$ & all
real $k_{x}$ \\ \hline
Always Cutoff & $\varepsilon _{y}\mu _{x}<0$ & $\mu _{x}/\mu _{z}>0$ & no
real $k_{x}$ \\ \hline
\end{tabular}%
\end{equation*}%
Note the analysis presented here is carried out at constant frequency, and
that the term \emph{cutoff} always refers to the transverse component of the
wave vector, $k_{x}$, not the frequency, $\omega $. \ Iso-frequency
contours, $\omega \left( \mathbf{k}\right) =const$, show the required
relationship between $k_{x}$ and $k_{z}$ for plane wave solutions (Fig. \ref%
{dispersion and refraction fig}). \ \ 

\begin{figure}[tbp]
\includegraphics[ width=3.18in, height=7.02in ]{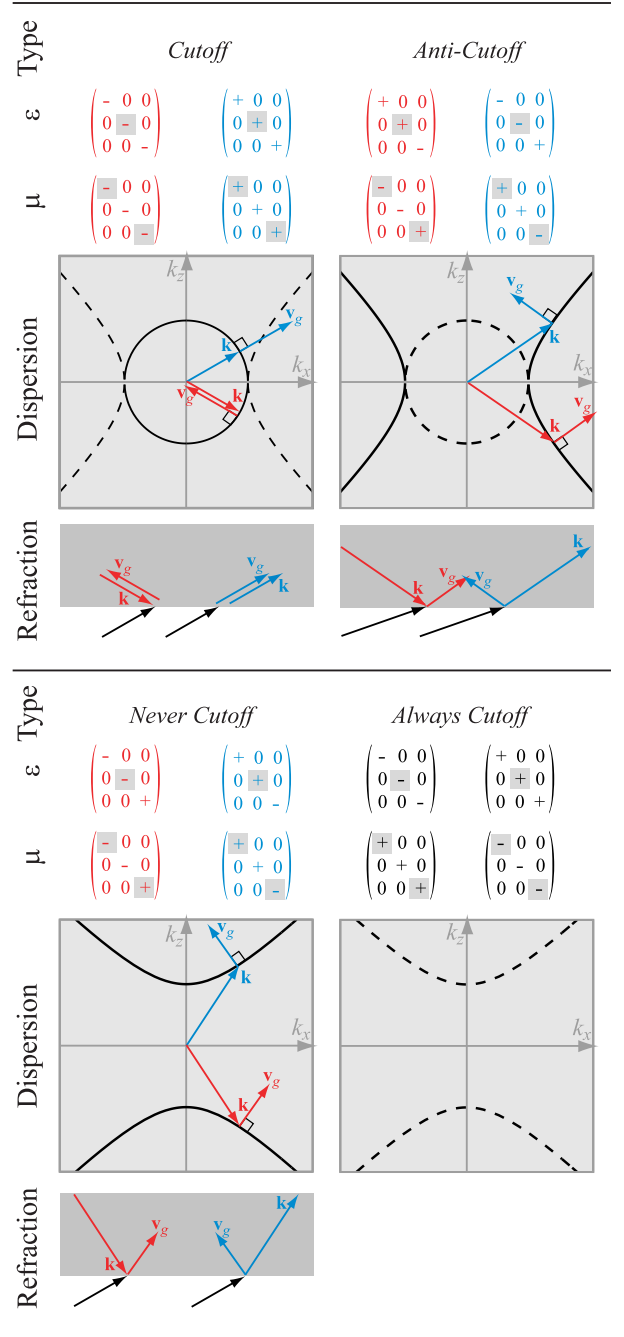}
\caption{Material property tensor forms, dispersion plot, and refraction
diagram for four classes of media. \ Each of these media has two sub-types:
one positive (blue) and one negative (red) refracting, with the exception
that \emph{always cutoff} media does not support propagation and refraction.
\ The dispersion plot shows the relationship between the components of the
wave vector at fixed frequency. $k_{x}$ (horizontal axis) is always real, $%
k_{z}$ (vertical axis) can be real (solid line) or imaginary (dashed line).
\ The same wave vector and group velocity vectors are shown in the
dispersion plot and the refraction diagram. \ $\mathbf{v}_{g}$ shows
direction only. The shaded diagonal tensor elements are responsible for the
shown behavior for electric $y$-polarization, the unshaded diagonal elements
for magnetic $y$-polarization.}
\label{dispersion and refraction fig}
\end{figure}

To understand wave refraction at an interface between vacuum and an
indefinite medium, we must first examine the general relationship between
the directions of energy and phase velocity for waves propagating within an
indefinite medium. We can accomplish this by calculating the group velocity, 
$\mathbf{v}_{g}\equiv \nabla _{\mathbf{k}}\omega \left( \mathbf{k}\right) $. 
$\mathbf{v}_{g}$ specifies the direction of energy flow for the plane wave,
and is not necessarily parallel to the wave vector. $\nabla _{\mathbf{k}%
}\omega \left( \mathbf{k}\right) $ must lie normal to the iso-frequency
contour, $\omega \left( \mathbf{k}\right) =$ constant, but may lie on either
side, as illustrated in Fig. \ref{dispersion and refraction fig}.\ To
determine the correct direction, we utilize the dispersion relation in Eq. (%
\ref{dispersion}) to calculate $\nabla _{\mathbf{k}}\omega \left( \mathbf{k}%
\right) $. \ Performing an implicit differentiation of Eq. (\ref{dispersion}%
) leads to a result for the gradient that does not require square root
branch selection, removing any sign confusion. \ To obtain physically
meaningful results, a causal, dispersive response function, $\xi \left(
\omega \right) $, must be used to represent the negative components of $%
\varepsilon $ and $\mu $, since these components are necessarily dispersive %
\cite{landau10}. \ The response function should assume the desired
(negative) value at the operating frequency, and satisfy the causality
requirement that, $\partial \left( \xi \omega \right) /\partial \omega \geq
1 $ \cite{landau10,smithkroll}. Combing this with the derivative of Eq. (\ref%
{dispersion}) determines which of the two possible normal directions
applies, without specifying a specific functional form for the response
function. Fig. \ref{dispersion and refraction fig}\ relates the direction of
the group velocity to a given material property tensor sign structure.

Having calculated the energy flow direction, we can determine the refraction
behavior of indefinite media by applying two rules: (1) the transverse
component of the wave vector, $k_{x}$, is conserved across the interface,
and (2) energy carried into the interface from free space must be carried
away from the interface inside the media; i.e., the normal component of the
group velocity, $v_{gz}$, must have the same sign on both sides of the
interface. \ Fig. \ref{dispersion and refraction fig} shows typical
refraction diagrams for the three types of media that support propagation.

Despite the interesting refraction properties associated with indefinite
media, a finite slab of indefinite material will in general present a
significant impedance mismatch to waves incident from vacuum. \ We can,
however, increase the transparency at a given frequency by combining a layer
of indefinite material with a second layer possessing the same dispersion
and impedance but opposite refractive index. Such a compensated bilayer will
have inter-layer relative impedance near unity, and will exhibit zero phase
shift from one side to the other, for all plane wave components.

To illustrate the possibilities associated with compensated bilayers of
indefinite media, we first recall that a motivating factor in the recent
metamaterials effort has been the prospect of near-field focusing. \ Pendry
showed theoretically that a planar slab with isotropic $\varepsilon =\mu =-1$%
, could act as a lens with resolution well beyond the diffraction limit. It
is difficult, however, to realize significant sub-wavelength resolution with
an isotropic negative index material, as the required exponential growth of
the large $k_{x}$ field components across the negative index lens leads to
extremely large field ratios\cite{ziolkowski}. Sensitivity to material loss
and other factors can significantly limit the sub-wavelength resolution.

\begin{figure}[tbp]
\includegraphics[ width=3.38in, height=1.48in ]{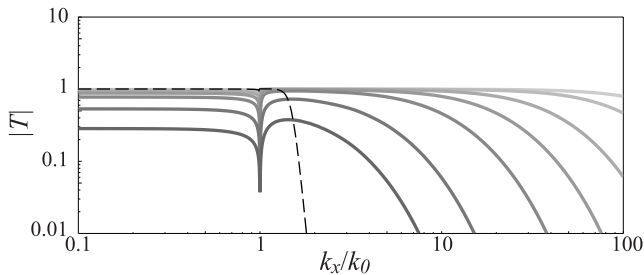}
\caption{Magnitude of the transfer function vs. transverse wave vector, $%
k_{x}$, for a bilayer composed of positive and negative refracting \emph{%
never cutoff} media. Material property elements are of unit magnitude and
layers of equal thickness, $L/\protect\lambda $ = 0.1, 0.2, 0.5, 1, 2. The
thinnest layer is the darkest curve. A realistic loss of $0.01i$ has been
added to each diagonal component of $\mathbf{\protect\varepsilon }$ and $%
\mathbf{\protect\mu }.$ For comparison, a single layer, isotropic near field
lens is shown dashed. \ The single layer has thickness, $\protect\lambda $,
and $\protect\varepsilon =\protect\mu =-1+0.001i$.}
\label{transfer function fig}
\end{figure}

By combining positive and negative refracting layers of \emph{never cutoff}
indefinite media, we can produce a compensated bilayer that accomplishes
near-field focusing in a similar manner to the perfect lens, but with
significant advantages. Fig. \ref{dispersion and refraction fig} indicates
that for the same incident plane wave, the $z$-component of the transmitted
wave vector is of opposite sign for these two materials. Combining
appropriate lengths of these materials results in a composite indefinite
medium with unit transfer function. We can see this quantitatively by
considering the general expression for the transfer function of a bilayer. 
\begin{equation}
T=8\left[ 
\begin{array}{c}
e^{i\left( \phi +\psi \right) }\left( 1-Z_{0}\right) \left( 1+Z_{1}\right)
\left( 1-Z_{2}\right) + \\ 
e^{i\left( \phi -\psi \right) }\left( 1-Z_{0}\right) \left( 1-Z_{1}\right)
\left( 1+Z_{2}\right) + \\ 
e^{i\left( -\phi +\psi \right) }\left( 1+Z_{0}\right) \left( 1-Z_{1}\right)
\left( 1-Z_{2}\right) + \\ 
e^{i\left( -\phi -\psi \right) }\left( 1+Z_{0}\right) \left( 1+Z_{1}\right)
\left( 1+Z_{2}\right)%
\end{array}%
\right] ^{-1}  \label{transfer}
\end{equation}%
The relative effective impedances are defined as 
\begin{equation}
Z_{0}=\frac{q_{z1}}{\mu _{x1}k_{z}},\;\;\;Z_{1}=\frac{\mu _{x1}}{\mu _{x2}}%
\frac{q_{z2}}{q_{z1}},\;\;\;Z_{2}=\mu _{x2}\frac{k_{z}}{q_{z2}}.
\label{impedances}
\end{equation}%
where $\mathbf{k}$, $\mathbf{q}_{_{1}}$ and $\mathbf{q}_{2}$ are the wave
vectors in vacuum and the first and second layers of the bilayer
respectively. \ The individual layer phase advance angles are defined as $%
\phi \equiv q_{z1}L_{1}$ and $\psi \equiv q_{z2}L_{2}$, where $L_{1}$ is the
thickness of the first layer and $L_{2}$ is the thickness of the second
layer. If the signs of $q_{z1}$\ and $q_{z2}$\ are opposite as mentioned
above, the phase advances across the two layers can be made equal and
opposite, $\phi +\psi =0$. \ If we further have the condition that the two
layers are impedance matched to each other, $Z_{1}=1$, then Eq. (\ref%
{transfer}), reduces to $T=1$. \ In the absence of loss, the material
properties can be chosen so that this occurs for all values of the
transverse wave vector, $k_{x}$. \ \ A transfer function with some realistic
loss added is shown in Fig \ref{transfer function fig}.

A proposed implementation is shown in Fig. \ref{implementation fig}. The
elements shown in the top and bottom of the figure will implement media that
focuses electric $y$-polarized and magnetic $y$-polarized waves
respectively. Combining the two structures results in a bilayer that focuses
both polarizations and is $x$-$y$ isotropic. The materials are formed from
split ring resonators and wires with numerically and experimentally
confirmed effective material properties \cite{shelby}. Each split ring
resonator orientation implements negative permeability along a single axis,
as does each wire orientation for negative permittivity.

While compensated bilayers of indefinite media exhibit reduced impedance
mismatch to free space and high transmission, uncompensated semi-infinite
indefinite media exhibit unique high reflection properties. \ Aside from 
\emph{cutoff} materials, all other classes of indefinite media have a
reflection coefficient amplitude near unity for incident propagating waves.
The phase of the reflection coefficient, however, varies, as illustrated
here for positive refracting \emph{anti-cutoff} media. The reflection
coefficient for electric $y$-polarization is given by 
\begin{equation}
\rho =\frac{\mu _{x}k_{z}-q_{z}}{\mu _{x}k_{z}+q_{z}}
\label{reflection coeff}
\end{equation}%
where $\mathbf{k}$ and $\mathbf{q}$ are the wave vectors in vacuum and the
media respectively. For unit magnitude \emph{anti-cutoff} material we have,
from Eq. (\ref{dispersion}), 
\begin{equation}
q_{z}^{2}=-\frac{\omega ^{2}}{c^{2}}+k_{x}^{2}=-k_{z}^{2}
\end{equation}%
Thus $q_{z}=\pm ik_{z}$. The correct sign for positive refracting media, $+$%
, is determined by the requirement that the fields must not diverge in the
domain of the solution. We then have 
\begin{equation}
\rho =\frac{1-i}{1+i}=-i
\end{equation}%
The magnitude of the reflection coefficient is unity, with a phase of -90%
${{}^\circ}$
for propagating modes of all incident angles. An electric dipole antenna
placed $\lambda /8$ away from the surface would thus be enhanced by
interaction with this ``mirror'' surface. Customized reflecting surfaces are
of practical interest, as they enhance the efficiency of nearby antennas,
while at the same time providing shielding \cite{yablonovitch,ma}. In this
case, the invariance of the reflection phase over all angles of incidence
may be advantageous. Furthermore, surface modes, which correspond to poles
in Eq. \ref{reflection coeff}, are not supported on this interface. Surface
modes require evanescent solutions on both sides of the interface. It is
possible, as shown in Fig. \ref{dispersion and refraction fig}, to have no
overlap of evanescent solutions (dashed line) between \emph{cutoff} and 
\emph{anti-cutoff} media. In many communications applications the energy
lost to the excitation of surface modes is undesirable, as it represents
loss of signal.

In conclusion, we have begun to explore the properties of media with
indefinite $\mathbf{\varepsilon }$ and $\mathbf{\mu }$ tensors. Indefinite
media are governed by hyperbolic dispersion relations, previously found in
more exotic situations, such as relativistic moving media \cite{kongBook}. \
Consideration of layered structures has led to useful and interesting
reflection and refraction behavior, including a new mechanism for
sub-diffraction focusing. We note that neither the analysis nor the
fabrication of these media is complicated, and thus anticipate other
researchers will quickly assimilate the principles and design structures
with unique and technologically relevant properties.

We thank Claudio Parazzoli (Phantom Works, Boeing) for motivating this work.
This work was supported by DARPA through grants from ONR (Contract No.
N00014-00-1-0632) and AFOSR (Contract No. 78535A/416250/440000) and a grant
from AFOSR (Contract Number F49620-01-1-0440).

\end{document}